\documentclass[traditabstract]{aa} 
\usepackage{longtable}
\usepackage{amsmath}
\usepackage{graphicx}
\usepackage{txfonts}
\usepackage{natbib}
\usepackage{verbatim}
\begin{document}
%
   \title{TMCalc - A fast code to derive Teff and [Fe/H] for FGK stars\thanks{Based on observations collected
       at La Silla Observatory, ESO, Chile, with the HARPS spectrograph
       at the 3.6-m telescope (072.C-0488(E)). The code is available for download at www.astro.up.pt/$\sim$sousasag/TMCalc}}

   \author{S. G. Sousa\inst{1,}\inst{2}
          \and
	  N. C. Santos\inst{1,}\inst{3,}
	  \and G. Israelian\inst{2,}\inst{4}
          }

	  \institute{Centro de Astrof\'isica, Universidade do Porto, Rua das Estrelas, 4150-762 Porto, Portugal
	  \and Instituto de Astrof\'isica de Canarias, 38200 La Laguna, Tenerife, Spain
	  \and Departamento de F\'isica e Astronomia, Faculdade de Ci\^encias, Universidade do Porto, Rua do Campo Alegre, 4169-007 Porto, Portugal
	  \and Departamento de Astrofisica, Universidade de La Laguna, E-38205 La Laguna, Tenerife, Spain
}

   \date{}

 
\abstract{
We present a new direct spectroscopic calibration for a fast estimation of the stellar metallicity [Fe/H]. This calibrations was computed using a large sample of 451 solar-type stars 
for which we have precise spectroscopic parameters derived from high quality spectra. The new [Fe/H] calibration is based on weak Fe I lines, which are expected to be less dependent on surface gravity and 
microturbulence, and require only a pre-determination of the effective temperature. This temperature can be obtained using a previously presented line-ratio calibration. We also present a simple code 
that uses the calibrations and procedures presented in these works to obtain both the effective temperature and the [Fe/H] estimate. The code, written in C, is freely available for the community and 
may be used as an extension of the ARES code. We test these calibrations for 582 independent FGK stars. We show that the code 
can be used as a precise and fast indicator of the spectroscopic temperature and metallicity for dwarf FKG stars with effective temperatures ranging from 4500 K to 6500 K 
and with [Fe/H] ranging from -0.8 dex to 0.4 dex. 
}

\keywords{Stars: fundamental parameters - Stars: abundances - Stars: statistics - Methods: data analysis }
\authorrunning{Sousa, S. G. et al.}
\titlerunning{TMCalc - a fast code to derive Teff and [Fe/H] for FGK stars}

\maketitle

\section{Introduction}

The derivation of stellar parameters is of extreme importance for several fields of astrophysics. Our knowledge of the fundamental parameters of 
stars, such as the mass and radius, depends directly on the precision that we can achieve when measuring the stellar atmosphere parameters. Parameters that we can 
directly derive from observations include the effective temperature, surface gravity, and metallicity.

The two major techniques that are normally used to determine these stellar parameters are photometry and spectroscopy. There are many 
calibrations to help us determine parameters based on the first technique. The IRFM method is one of these methods, that nowadays, is very commonly used 
because of its reliability in determining the effective temperature of a wide range of stellar 
types \citep[][]{Blackwell-1977, Blackwell-1979, Blackwell-1980, Bell-1989, GonzHern-2009, Ramirez-Melendez-2005, Alonso-1996a, Casagrande-2006, Casagrande-2010}. Other parameters such as the metal 
content can also be derived from photometric calibrations \citep[][]{Nordstrom-2004, Schuster-1989}. These parameters can then be used to derive more fundamental parameters, 
such as mass and radius using calibrations similar to those described in \citet[][]{Torres-2010}.

The determination of the spectroscopic parameters is not straightforward. A careful analysis is necessary for each stellar spectrum. This can be 
very time-consuming if you need precise parameters and chemical abundances for several elements. One of the most difficult aspects of a spectroscopic 
analysis is, undoubtably, the continuum determination. It can be very difficult to derive correct position of the continuum, especially when dealing with 
poor quality spectra (considering data of low signal-to-noise (S/N) and/or low spectral resolution). Moreover, this difficulty in determining 
the continuum position also depends on the spectral type of a star. This is because cooler stars tend to have more spectral lines, increasing 
the amount of blending and leading to the ``dilution'' of the true continuum. The same happens for different wavelength regions of 
a given stellar spectrum where the line density strongly increases as you move to shorter wavelengths. This problem can be strongly reduced using automatic 
tools that are consistent in determining the continuum position, eliminating the errors/offsets, caused by a subjective determination, that are present when using interactive tools.

Another problem that has an important impact on the derivation of spectroscopic stellar parameters is the selection of the lines and their 
respective atomic parameters. There are two typical choices made by spectroscopists. On the one hand, one can choose to use for each line the atomic 
parameters defined by a laboratory analysis. This method can have errors as large as 10-20 \%. On the other hand, one can adopt 
new atomic-parameter values determined based on the observed lines in the spectra of a reference star (typically the Sun) and then assume 
its ``well'' known spectroscopic parameters. This second option allows us to perform a differential analyses using another star as a reference, and, 
therefore, when the Sun is used as a standard candle, it is very suitable for solar-type stars (FGK). However, the method becomes imprecise for sufficiently cool and hot stars.

This second option, which is usually referred to in the literature as a differential spectroscopic analysis, was combined previously with automatic codes such as 
ARES \citep{Sousa-2007} applied in our previous studies \citep[][]{Sousa-2008, Sousa-2010, Sousa-2011}. This allowed us to derive in 
a systematic and homogeneous way spectroscopic stellar parameters for more than 1000 FGK stars. These were then used to derive abundances for a large number
of elements \citep[][]{Neves-2009, Adibekyan-2011}. In all these works, the derived stellar 
parameters were proven to be compatible with others derived in different works using a range of independent methods.

In this work, we make use once again of the data of 451 stars presented in \citet[][]{Sousa-2008}, which consist of very-high quality spectra of both 
high resolution and high S/N, to perform a new direct spectroscopic calibration of the stellar metallicity [Fe/H]. The calibration is based on weak Fe I lines 
that depend mainly on temperature and iron abundance and are less dependent on other parameters such as the surface gravity and microturbulence. 
Therefore, it only requires a pre-determination of the effective temperature. For the determination of the temperature, we can use the line-ratio 
calibration presented in \citet[][]{Sousa-2010} and combine both calibrations to build a simple and fast code that allows a precise estimation of 
both the effective temperature and [Fe/H].

In Section 2 we recall and discuss the main features of the line-ratio calibration code, presenting some tests done in previous studies to demonstrate 
its consistency in the particular parameter space of the calibration. In Section 3, we define the new [Fe/H] spectroscopic calibration and explain how 
we derive it. We also present a simple procedure to derive the final value of [Fe/H] obtained from all the individual line calibrations. In Section 4, we show some 
simple tests performed on both the calibration sample and an independent large sample of solar-type stars. In the final Section 5, we summarize this work.

\section{Temperature based on a line-ratio calibration}

The code presented in this work is inspired by a previously published effective-temperature calibration based on line ratios of several spectral lines of different 
elements \citep[][]{Sousa-2010}. This calibration was presented as an excellent tool for determining automatically and quickly a spectroscopic 
effective temperature, and can be easily used to confirm a spectroscopic temperature determined using the ``standard'' 
procedures \citep[see also][]{Gray-1991, Gray-1994, Gray-2001,Gray-2004, Kov-2003}. This was presented as a possible 
extension of the ARES code, which can automatically measure the equivalent widths (EWs) of weak absorption spectral lines \citep[][]{Sousa-2007}. 
\citet[][]{Desidera-2011} showed that this line-ratio calibration can return good estimates of the temperature, even for solar-type stars up to 
relatively high rotation rates (up to 18 Km/s). This excedes the abilities of typical EW methods owing to the increase in the number of blends with higher 
rotation of the stars.

\begin{figure}[!t]
\centering
\includegraphics[width=8cm]{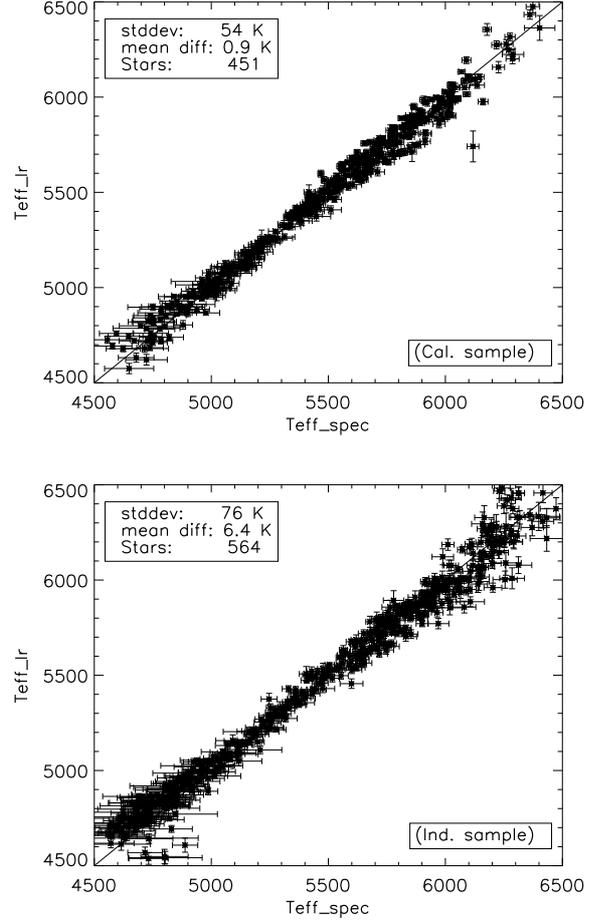}
\caption[]{In the top panel, we show the comparison between the effective temperatures derived from our standard spectroscopy analysis and the ones 
derived using the line-ratio code for the sample used to derive the calibration (labelled Cal. Sample in the plot adapted from \citet[][]{Sousa-2010}). In the bottom plot, 
we show a similar plot but this time for the sample of stars presented in \citet[][]{Sousa-2011}. This corresponds to a sample that is independent 
of the derived calibration. In both panels, we present the mean difference and the standard deviation for the comparisons.}
\label{fig1}
\end{figure}

In Figure \ref{fig1}, we compare the temperatures derived with the line-ratio calibration of \citet[][]{Sousa-2010} with those derived 
using the standard spectroscopic procedure \citep[][]{Sousa-2008, Sousa-2011}. The top plot shows this comparison for 
the sample of stars used to calibrate the line ratios. As expected, the consistency is very good. This plot is important to illustrating 
the typical dispersion that we obtain using the line-ratio calibration. 

In the bottom plot of the same figure, we show the same comparison for an ``independent'' sample of stars (independent in the sense that these stars were not used to compute 
the calibration of each line ratio). It is clear that the comparison is also consistent. Both plots were presented in previous works, and they 
are presented here to recall that the temperature inferred from the line ratio is consistent with our standard spectroscopic method, within the ranges 
defined for the calibration. Therefore, we can use the line-ratio calibration in order to determine the temperature of a star, and then examine the strength of the 
iron absorption-lines and quickly find a calibration of [Fe/H].

\section{[Fe/H] calibration}

\subsection{Selection of lines}

To obtain a calibration that help us determine [Fe/H], we need to compile a list of iron lines that are observed for solar-type stars. 
For this step, we started with the ``stable'' line-list presented in \citet[][]{Sousa-2008}, which is a list of iron lines chosen to ensure that they 
are appropriate for an automatic determination of EWs with ARES. The next step was to select lines that should be independent of both surface gravity 
and microturbulence. In other words we needed at this point, to select lines that depend mostly on temperature and [Fe/H].


We can easily ensure the independence of the calibration of the surface gravity by selecting only the iron lines that corresponds to the neutral state (Fe I). This is because 
the ionized iron lines (e.g. Fe II) depend strongly on the surface gravity, in contrast to the neutral iron lines \citep[][]{Gray-1992}. 
We therefore take out all the Fe II lines from the original list.

Finally, the microturbulence parameter may have a strong impact on the strength of a spectral line, particulary for stronger lines. Weak lines are known to be reasonably independent of microturbulence.


One problem that we face in defining our linelist is that the strength of each individual line depends strongly on the effective temperature and [Fe/H]. We can therefore only eliminate 
a line after its line strengh has been measured and checked to be in the regime where it is independent of microturbulence. Therefore, for this work we 
consider our cut at 75 m\AA. This value seems to be reasonable taking into account not only the microturbulence dependence, but also considering that lower values will 
strongly reduce the number of lines acceptable for the calibrations.

We also wish to note that there is a known dependence of the microturbulence on the temperature \citep[e.g.][]{Pilachowski-1996, For-2010}. Therefore, the microturbulence 
dependence seen for the stronger lines may be removed significantly by the temperature fitting in our calibrations. Together with the restrictions that we impose on the line selection 
we can strongly eliminate any microturbulance dependence from our calibrations.

\begin{figure}[!t]
\centering
\includegraphics[width=8cm]{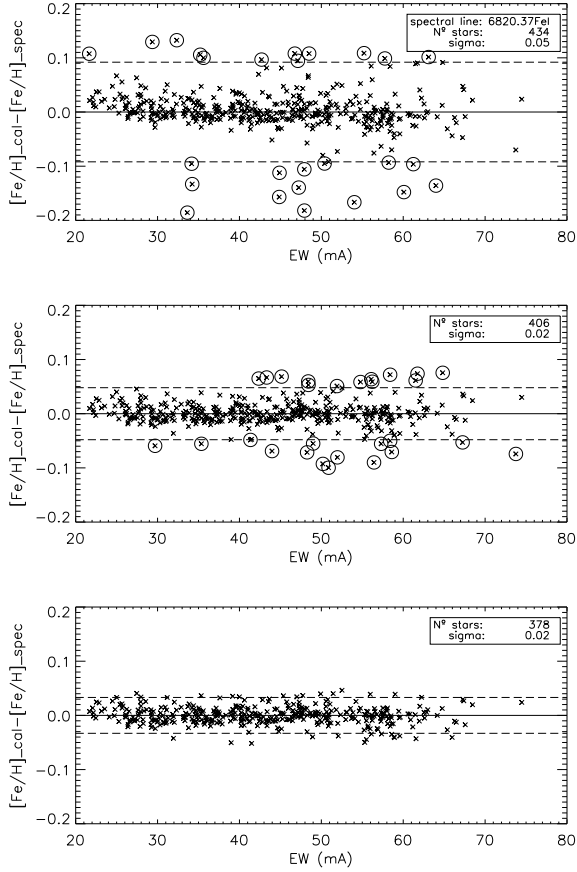}
\caption[]{The sequence of plot from top to bottom shows the removal of outliers for the iron line at 6820.37\ \AA.}
\label{fig2}
\end{figure}

\subsection{The [Fe/H] calibration}

To derive the calibration for [Fe/H], we again used the sample of stars presented before for the temperature line-ratio calibration. This sample is composed of
451 solar-type stars that were all observed with HARPS to obtain high-resolution spectra (R $\sim$ 110000). These stars were part of the HARPS survey for extra-solar planets \citep[][]{Mayor-2011}. 
The S/N for this sample varies from 70 to 2000, with 90\% of the 
spectra having a S/N higher than 200. It is a very well-established sample with homogeneous spectroscopic parameters that have been compared with other 
independent methods.

To ensure consistency with the previous line-ratio calibration, we used the same EW measurements.

For each line, a calibration was computed following the equation:

\begin{equation}
\begin{split}
EW = & C_0 + C_1 *[Fe/H]  + C_2 * T_{eff} + \\
     & C_3 * [Fe/H]^2 + C_4 * T_{eff} ^ 2 + \\
     & C_5 * [Fe/H] * T_{eff}.
\end{split}
\end{equation}

We choose to use this equation in order to have an easy (and quickly attainable) representation of the physical dependence between the line strength, the effective temperature, and the iron abundance. 
To extract the iron abundance from this equation, we perform a simple invertion of $[Fe/H]$.

In Fig. \ref{fig2}, we show the results of an iterative process for the determination of the calibration coefficients for the iron line at 6820.37 \AA. In this process, we first select the lines (stars) 
accordingly to the discussion in Sect. 3.1, i.e. removing all points whith EWs larger than 75 m\AA\ (each point is the EW for a star in the sample). At this step, we also choose to 
remove the lines with EWs smaller than 20 m\AA. This is because for these small lines we have a larger relative error in the EW. We then fit the remaining points with the equation and 
perform the first outlier removal, choosing a 2 $\sigma$ threshold. We repeat the process one more time to eliminate extra outliers. These outliers are mainly due to poor automatic EW measurements that 
are typically related to either a bad continuum determination or strongly blended lines. The final fitted coefficients are kept, together with the final number of stars used, the slope of the direct comparison 
between the calibrated and the spectroscopic [Fe/H], and the respective standard deviation 
of the fit. 


This procedure was performed for all the lines in the initial iron line-list. Finally, we made a careful selection of the lines, considering the results 
of each individual fit. This selection was performed to ensure that each calibration satisfies the following conditions:


\begin{enumerate}
 \item The direct comparison between the calibrated and the spectroscopic [Fe/H] has a slope within 3\% of the identity line.
 \item The standard deviation of the individual line calibration is less than 0.04 dex.
\end{enumerate}

The values used in these two conditions were chosen to keep a significantly large number of lines in order to increase the statistical meaning of the final metallicity derivation. In this case we 
want to note we choose not to neglect lines that have a given minimum of the final number of stars used for the individual fit. The reason for this is that we 
remove lines of the fit accordingly to their strength, which means that depending on the temperature and metallicity of a star, a given line will be either stronger or weaker in 
specific space regions of these two parameters. Therefore, the only concern that we have in these cases is to ensure that we take into account that each individual calibration is only valid 
within each individual parameter space.


Table \ref{tab2} lists a sample of the total of 149 lines that have passed this process and can be used for our [Fe/H] determination. The full table is available in its electronic format.

%

\subsection{[Fe/H] estimation from the line calibrations}

From the calibrated lines, we can now derive a final value for the global [Fe/H] for a given star. The procedure that we propose here is the same as that 
presented for the temperature determination based on line ratios \citep[][]{Sousa-2010}. We summarize this procedure in the following items:

\addtocounter{table}{1}

\begin{itemize}
 \item First we compute the [Fe/H] determination using each calibrated line from Table \ref{tab2}.
 \item Secondly, we select the calibrated lines accordingly to the limits of each individual calibration. In this step, we choose to increase the limits by 
100K in both directions. The errors coming from the temperature line-ratio calibration are of this order of magnitude and therefore we wish to guarantee that we do not remove 
lines that can give a reasonable estimation of [Fe/H].
 \item Finally, we compute the weighted average of the [Fe/H] results, considering the standard deviation of each individual calibration.
\end{itemize}

This procedure is repeated twice with a 2 $\sigma$ outlier removal, eliminating in this process the EWs that were not (for any reason) performed correctly.

\subsection{Errors}

The easiest error estimate that we can extract from our procedure is to assume the dispersion in the values given by all the individual 
calibrations. If one considers that each individual calibration is independent of all the others, we can divide the dispersion by the square root of the 
final number of individual calibrations used.

It is wise, however, to include the error in the temperature. To do this, we choose to use a straightforward estimation of this error, which is to derive the 
error in [Fe/H] using the limits given by the temperature error (1-$\sigma$). The final error is obtained by evaluating the quadratic sum of the two sources of errors.

\subsection{Using TMCalc to estimate temperature and [Fe/H] for solar-type stars}

The calibration presented here is only useful when you have a temperature estimation. We therefore developed a free code, implemented in C language, that combines both the line 
ratio calibration and the [Fe/H] calibration presented here. The code is available online at the ARES web page\footnote{http://www.astro.up.pt/$\sim$sousasag/ares)}. 
The code comes with a simple shell script ``TMCalc'', which can be used as the driver to run the C code with an easy shell-command line. This makes the code ready to be included 
in any kind of a spectral analysis pipeline. The only requirement is to have the spectrum (e.g. coming from the pipeline) in a format ready to be used with ARES. 

\section{Testing the code}

\subsection{The calibration sample}

\begin{figure}[!t]
\centering
\includegraphics[width=8cm]{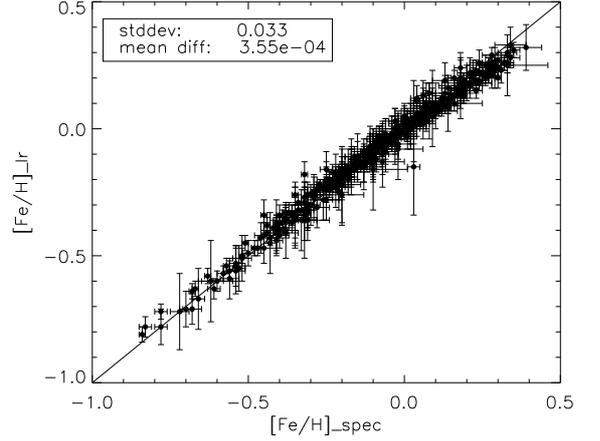}
\caption[]{Direct comparison between the calibrated [Fe/H] and the spectroscopic [Fe/H] for the sample used to compute the calibrations.}
\label{fig3}
\end{figure}

Figure \ref{fig3} shows the results obtained when applying the code to the calibration sample. Here we compare the calibrated [Fe/H] against the spectroscopic [Fe/H]. 
The result is consistent, thus can be expected since in this case the sample is the same as that used to compute the line calibrations.

\begin{figure}[t]
\centering
\includegraphics[width=8cm]{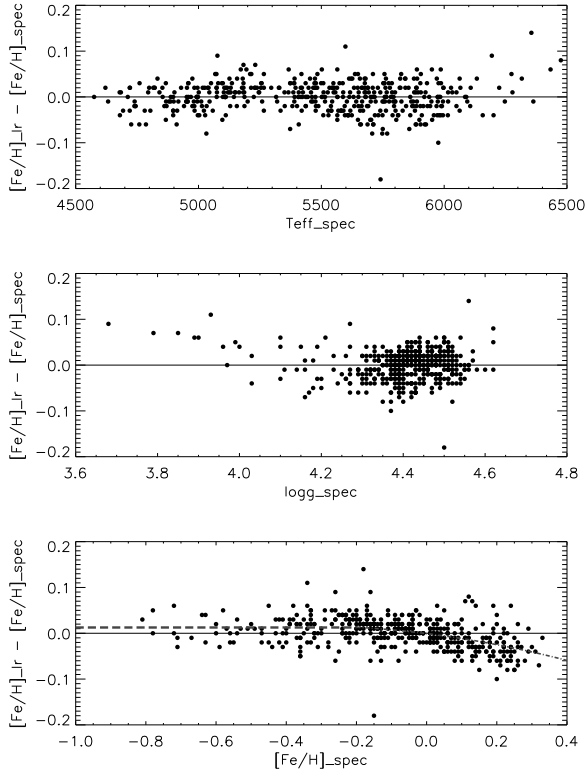}
\caption[]{Difference between the calibrated [Fe/H] and the spectroscopic [Fe/H] and its dependence on the other spectroscopic stellar 
parameters such as the effective temperature (top panel), the surface gravity (middle panel), and the spectroscopic [Fe/H] (bottom panel). 
The dashed and dashed pointed lines are fits to the date described in detail in the text.}
\label{fig4}
\end{figure}

In Fig. \ref{fig4}, we try to identify any dependences of the value of [Fe/H] derived from the calibration on other spectroscopic parameters such as 
temperature, surface gravity, and spectroscopic [Fe/H]. It is possible to see from this figure the ranges of spectroscopic 
parameters for which this calibration is valid. It is clear that these come directly from the star properties in the sample, which is typically composed of 
solar-type stars with effective temperatures ranging between 4556 K and 6403 K, and with surface gravities typical of dwarf stars 
and a few sub-giants ([3.68, 4.62]). All these stars have metallicities of around solar, ranging from -0.84 dex to 0.39 dex.

It is clear from Fig. \ref{fig4} that there is no clear trend, except for the spectroscopic [Fe/H] itself. An offset is clearly visible for the metal-rich stars, but 
the maximum differences remain within 0.1 dex, which can nevertheless be significant. This offset is a result of some extra dependence that the individual line calibrations 
were unable to extract, or is merely an artifact of the fit, since the offset is only observed in the upper limit to [Fe/H]. 
Instead of applying a higher order polynomial for each individual calibration and trying to eliminate this offset, we choose to perform a simpler and more direct correction. 
From the figure, we can see that we can fit the points with a simple horizontal line for [Fe/H] $< -0.2$ (dashed line). This corresponds to a calibrated [Fe/H] $\sim -0.187$, given 
that the horizontal line has a value of $\sim 0.013$. For higher [Fe/H] values, we use a second-order polynomial to consistently fit the offset (dashed-point line). We therefore use the 
following equations to perform the correction for the calibrated [Fe/H] in these two regimes:


\begin{equation}
\begin{split}
[Fe/H]_{cor}=&[Fe/H]_{cal} - (0.013)\\
	      & \mbox{ if  } ( [Fe/H]_{cal} < (-0.187) ),\\
\\
[Fe/H]_{cor}=&[Fe/H]_{cal} - (-0.000726960 \\
	      &-0.0950966 * [Fe/H]_{cal} \\
	      &-0.128329* [Fe/H]_{cal}^2 \\ 
	      & \mbox{ if  } ( [Fe/H]_{cal} \ge (-0.187) ).\\
\end{split}
\end{equation}

This equation is applied at the end of the procedure to correct the offset that can be seen in Figure \ref{fig4}.

\begin{figure}[h]
\centering
\includegraphics[width=8cm]{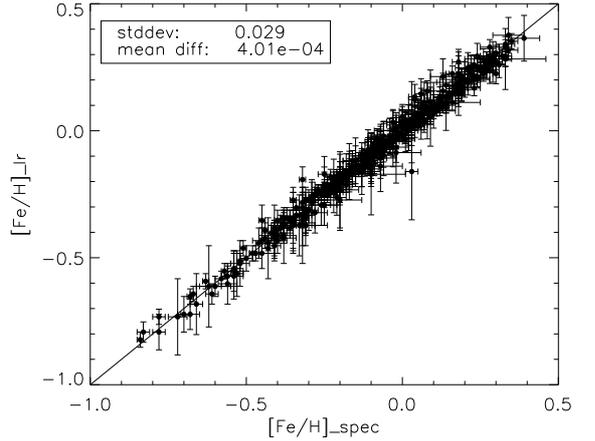}
\caption[]{Direct comparison between the corrected calibrated [Fe/H] and the spectroscopic [Fe/H] for the sample used to compute the calibrations.}
\label{fig5}
\end{figure}

In Figure \ref{fig5}, we show the effect of the correction for the sample used for the calibration. The dispersion in slightly smaller and the offset is 
close to zero as expected. However, this is still the sample of stars used to derive the calibration, hence the consistency of this result is 
expected. What is interesting to show at this stage is the ``small'' dispersion in the evaluated [Fe/H], which is typically around $\sim$ 0.03 dex. This 
value can be used as a reference to indicate the quality of the final corrected calibration.

\subsection{Constrains on line strenghts}

As discussed before, we choose lines of a specific range of strengths: weak lines in order to avoid 
any dependence on the microturbulence, but not those that are weak to avoid the intrinsic errors in the measurements of very weak lines. The reader may be concerned about 
these restrictions and the possible systematics that they can produce on the described calibrations. For instance, for low temperature stars, these restrictions 
will tend to eliminate the stronger lines in the metal-rich stars, while in hotter stars, different lines (that are weaker) will be eliminated for the metal-poor stars owing to the 
poorer constraints. Even without our poorer constraints, the iron lines tend to disappear from the spectrum in these cases, since they are metal-poor. Therefore, for 
different types of stars, we use different sets of lines, which could introduce systematic errors into the calibrations.

\begin{figure*}[!t]
\centering
\includegraphics[width=16cm]{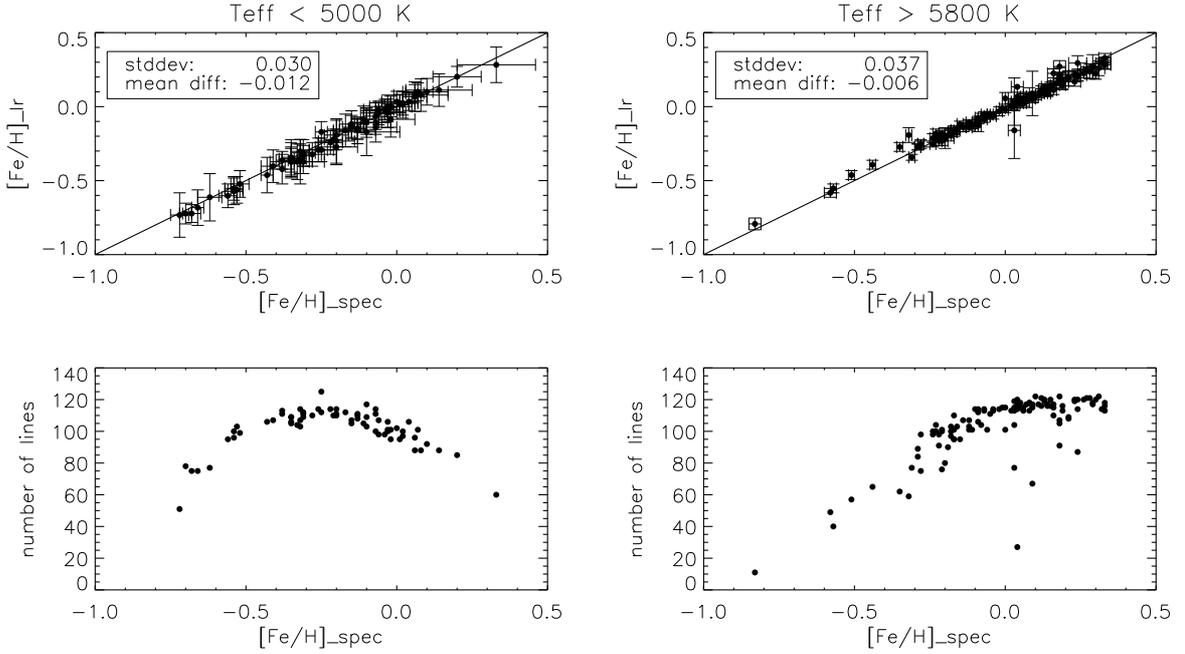}
\caption[]{Direct comparison between the corrected calibrated [Fe/H] and the spectroscopic [Fe/H] for the sample used to compute the calibrations 
for two different temperature regimes: cool stars (Top left panel) and hotter stars (top right panel). In the bottom panel, we present the number of 
individual calibrations used in each star.}
\label{fig_diff_teff}
\end{figure*}


Figure \ref{fig_diff_teff} tries to identify any of the described possible systematic errors. In the top panels of this figure, we show the same kind of direct comparison between the corrected 
calibrated [Fe/H] and the spectroscopic [Fe/H], but divide the sample into different temperature regimes. On the left we see the lower temperature stars ($T_{eff} <$ 5000 K), and on the right 
the hotter stars ($T_{eff} >$ 5800 K). The comparisons remains consistent with a small dispersion and no visible trend for the different temperature regimes. In the bottom panels, we present 
the number of individual calibrations used in each star. In these plots the decrease in the number of lines used is clear for the metal-poor stars in both temperature regimes. For the cooler 
stars, the reduction in the number of lines is also clear for the more metallic stars. This is the effect that we described 
before and is a result of neglecting the stronger lines to eliminate the microturbulence dependence. Although the number of lines is smaller owing to the discussed systematics, it is clear 
from the comparison that the results are consistent, and the selection of the lines using the indicated constraints on the line strength seem to produce good results.

\subsection{Large independent sample for direct comparison}

We only showed the test on a very well-defined sample, composed of stars with data of high resolution and high S/N, which was used to define the calibration.
At this point, we wish to show an independent test, in the sense that we use a different sample of stars that was not used to compute the calibration. This ``independent'' sample 
is composed of a total of 582 FGK stars for which the parameters were determined using the same homogeneous spectroscopic method. The details of these determinations can be seen in 
\citet[][]{Sousa-2011}. The main difference regards the spectral quality of this sample: this sample typically has a lower S/N than the calibration sample. 75\% of the stars have S/N 
lower than 200, while 90\% of the stars in the calibration sample have data of S/N higher than 200.

\begin{figure}[!t]
\centering
\includegraphics[width=8cm]{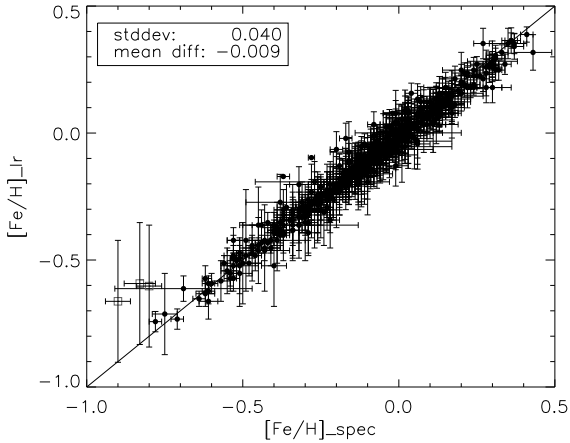}
\caption[]{Direct comparison of the corrected calibrated [Fe/H] and the spectroscopic [Fe/H] for an independent sample of solar-type stars. The three grey squares indicate 
stars for which the code obtained results outside the calibrated [Fe/H].}
\label{fig6}
\end{figure}

In Figure \ref{fig6}, we compare the standard spectroscopic derivation result with the calibration presented here for this ``independent'' large sample of stars.

The result of this direct comparison is very good. The mean difference is close to zero, indicating that there is no clear offset in the 
determined [Fe/H]. However, the dispersion ($\sim$0.04 dex) is larger than for the previous comparisons. This is expected because this sample 
has typically a lower S/N. In addition, this large ``independent'' sample has a few stars that lie outside the limits of the calibration. Amoung the 582 stars 
that belong to this sample, the code was able to derive values for a total of 556 stars. Amoung these stars, there are still a few that are outside the limits of the calibration. 
This can be clearly seen in the figure at lower metallicities where we have at least three stars with metallicities lower than -0.8 dex (plotted as grey squares in the Figure). 
The error bars for these three stars are large compared to those for the rest of the stars. This is due to the far smaller number of iron lines (about ten) used in the estimation of the calibrated [Fe/H]. 
The number of lines and their respective errors derived from the procedure previously described can be used to reliably check whether the star is within the limits of the parameter space of the calibration.

\begin{figure}[!t]
\centering
\includegraphics[width=8cm]{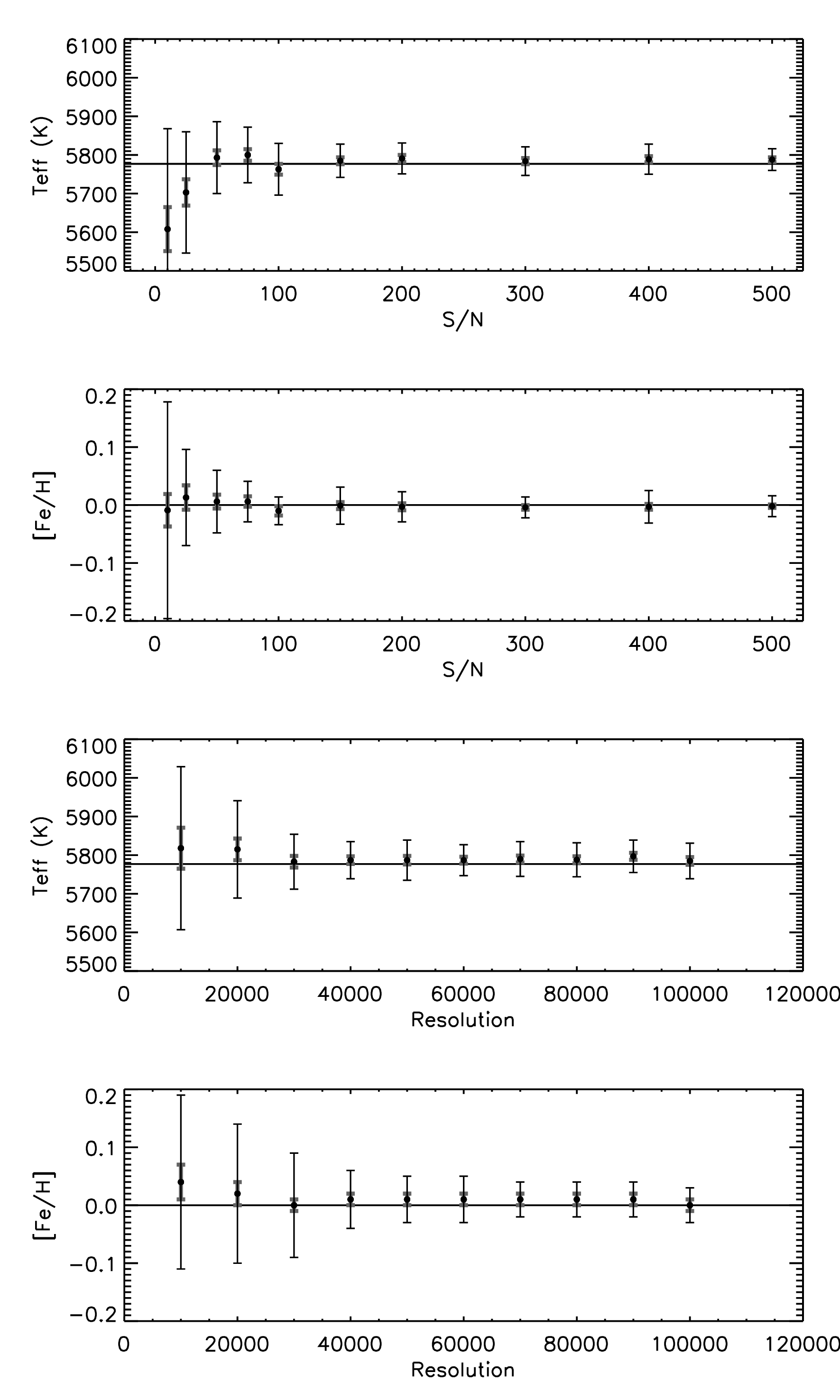}
\caption[]{Temperature and [Fe/H] determined using our calibrations for the simulated solar spectrum with different sets of resolutions and S/N. We show two different error estimations, 
the thin black error bars representing only the dispersion in the individual calibrations, and the thick grey error bar where we also take into account the number of individual calibrations used. 
See text for details.}
\label{fig7}
\end{figure}

After this has been noted, the remaining results are very consistent and reveal that this calibration, both in temperature and [Fe/H], is precise enough to be 
used within its limits, to efficiently derive these two atmosphere parameters, either as a first approximation or as a check of other standard spectroscopic 
determinations.

\subsection{Spectral resolution and S/N}

To evaluate how this procedure works for different spectral resolutions and different values of S/N, we computed several solar spectra, first with different levels of S/N and then with different 
levels of resolution. The original spectrum was obtained using the Kurucz Solar Atlas. The noise was created using a Gaussian distribution and we introduced the artificial instrumental resolution 
using the ``rotin3'' routine in SYNSPEC\footnote{http://tlusty.gsfc.nasa.gov/index.html} \citep[][]{Hubeny-1994}. We first created ten different S/N levels to be included in the original 
spectrum (10, 25, 50, 75, 100, 150, 200, 300, 400, and 500). We then created ten addicional different spectra with resolutions ranging from 10000 to 100000, corresponding to 
FWHM of 0.550\ \AA\ and 0.055\ \AA\  at the central wavelength 5500\ \AA. For this spectra, we included the noise to simulate 
spectra with S/N $\sim 200$, which more closely represent the real spectra. We generated the spectra in the region [4000\ \AA\ - 7000\ \AA\ ] in order to cover our linelist.

For the spectra of different resolutions, the EW measurements were all performed with ARES using the same input parameters (smoothder=4, space=3, rejt=0.994, lineresol=0.07), except for the 
spectrum with the lowest resolution for which we adapted a ``smoothder'' parameter of 12. For the different S/N spectra, the ARES parameter ``rejt'' changed accordingly to the S/N of the 
spectra, using the recommended values discussed in \citet[][]{Sousa-2011}. The results can be seen in Figure \ref{fig7}. The two top panels shows the results for the solar spectra of different S/N, 
the two bottom plots for the solar spectra of different resolution. We also present two of the errors estimations given by the code. The smaller and thicker grey error-bars represent the errors 
obtained by considering the dispersion in the individual calibration and the number of independent lines used. The thin black error bars presented in this figure are the ones obtained using 
only the dispersion in the individual calibrations, ignoring the number of calibrations used. The results are consistent even for the lower resolutions, where the strong 
increase in the errors originating directly from dispersion in the independent calibrations can be clearly seen. This proves that the calibrations presented in this work can be safely used for 
lower resolution spectra. In terms of the S/N, the results are also quite consistent down to S/N $\sim$ 25. For this low S/N, the temperature estimation starts to deviate from the expected one, 
although the errors are large and greater than the offset. We note that for the [Fe/H] even for very low S/N spectra the values are very close to the expected [Fe/H]$=$0.

\section{Summary}

We have presented a new spectroscopic line calibration to efficiently estimate the stellar [Fe/H]. This calibration depends on both the temperature and the 
strength of the iron lines. We make available a free code that combined with ARES and a previous temperature line-ratio calibration allows us to estimate both the 
spectroscopic stellar effective-temperature and the [Fe/H].
We tested this code with a large sample of solar-type stars and confirmed that these calibrations are consistent within the parameter space defined 
for the calibrations. This code can easily be applied to a spectroscopic data-analysis pipeline in order to quickly obtain precise estimations of 
these important spectroscopic parameters.

These calibrations should not replace the more precise standard spectroscopic analyses that are typically more time-consuming. These standard methods should 
definitely be used if one whishes to study stars individually. However, the tool that we present in this work can be very useful for determining precise parameters for large amounts 
of data. Typical examples are the data sets of spectroscopic surveys, which are very large and for which people normally search for statistical trends among the propreties of 
the many survey targets.

\begin{acknowledgements}
S.G.S acknowledges the support from the Funda\c{c}\~ao para a Ci\^encia e Tecnologia (Portugal) in the form of a grants SFRH/BPD/47611/2008. NCS thanks for the 
support by the European Research Council/European Community under the FP7 through a Starting Grant. We also acknowledge support from FCT and FSE/POPH in the form of grants 
reference PTDC/CTE-AST/098528/2008, PTDC/CTE-AST/098754/2008, and PTDC/CTE-AST/098604/2008.

\end{acknowledgements}

\bibliographystyle{/home/sousasag/posdoc/mypapers/aa-package/bibtex/aa}
\bibliography{sousa_bibliography}

\longtab{2}{

\tiny

\begin{longtable}{c|ccc|rrrrrr|cc|cc}
\caption{\label{tab2} [Fe/H] line calibration table}\\
\hline\hline
$id$ & \textit{stddev} & \textit{ns} & $\lambda$ & $ c_0$ & $c_1$ & $c_2$ & $c_3$ & $c_4$ &  $c_5$ & $T_{eff}^{min}$ & $T_{eff}^{max}$ &  \textit{fehmin} & \textit{fehmax}  $\dagger$\\

\hline
\endfirsthead
\caption{continued.}\\
\hline\hline
$id$ & \textit{stddev} & \textit{ns} & $\lambda$ & $ c_0$ & $c_1$ & $c_2$ & $c_3$ & $c_4$ &  $c_5$ & $T_{eff}^{min}$ & $T_{eff}^{max}$ &  \textit{fehmin} & \textit{fehmax}  $\dagger$\\
\hline
\endhead
\hline
\endfoot
   1 & 0.038 &  296 & 6842.04 & -2.648e+02 &  1.116e+01 &  1.196e-01 &  1.559e+01 & -1.201e-05 &  5.558e-03 & 4578 & 6276 &  -0.45 &   0.39\\
   2 & 0.015 &  386 & 4523.40 &  3.340e+01 &  9.804e+00 &  3.054e-02 & -6.349e-01 & -4.963e-06 &  6.433e-03 & 4624 & 6403 &  -0.78 &   0.39\\
   3 & 0.022 &  227 & 4537.67 &  2.850e+02 &  1.119e+02 & -7.090e-02 &  1.515e+01 &  4.253e-06 & -1.374e-02 & 4578 & 6026 &  -0.72 &   0.39\\
   4 & 0.014 &  322 & 4551.65 & -7.152e+01 &  3.406e+01 &  5.782e-02 &  5.992e+00 & -7.036e-06 &  1.309e-03 & 4624 & 6287 &  -0.66 &   0.39\\
   5 & 0.030 &  339 & 4554.46 &  8.308e+02 &  1.017e+02 & -2.340e-01 & -5.797e+00 &  1.677e-05 & -1.028e-02 & 4649 & 6403 &  -0.72 &   0.39\\
   6 & 0.032 &  362 & 4561.41 &  2.503e+01 &  9.022e+01 &  2.946e-02 &  1.166e+01 & -4.717e-06 & -6.692e-03 & 4556 & 6374 &  -0.72 &   0.39\\
   7 & 0.015 &  396 & 4566.52 &  2.469e+01 &  4.034e+01 &  3.551e-02 &  6.998e+00 & -5.498e-06 &  1.472e-03 & 4624 & 6374 &  -0.78 &   0.39\\
   8 & 0.037 &  399 & 4574.22 &  1.069e+02 &  1.187e+02 &  1.120e-02 &  1.076e+01 & -3.925e-06 & -1.193e-02 & 4556 & 6361 &  -0.84 &   0.39\\
   9 & 0.035 &  341 & 4574.72 & -1.094e+02 & -4.674e+01 &  8.927e-02 & -5.797e+00 & -1.049e-05 &  1.508e-02 & 4679 & 6374 &  -0.84 &   0.35\\
  10 & 0.029 &  326 & 4579.33 & -1.005e+01 &  2.901e+01 &  4.280e-02 &  3.486e+00 & -6.285e-06 &  2.275e-03 & 4578 & 6276 &  -0.72 &   0.35\\
  11 & 0.018 &  348 & 4593.53 & -1.559e+02 &  4.349e+01 &  8.990e-02 &  1.158e+01 & -1.002e-05 &  1.026e-04 & 4578 & 6287 &  -0.66 &   0.39\\
  12 & 0.016 &  377 & 4596.41 & -6.229e+01 &  4.851e+01 &  6.032e-02 &  1.230e+01 & -7.593e-06 & -2.193e-04 & 4624 & 6361 &  -0.72 &   0.39\\
  13 & 0.029 &  168 & 4602.00 &  1.424e+02 & -6.960e+01 & -6.436e-04 & -9.508e+00 & -2.015e-06 &  1.797e-02 & 5023 & 6361 &  -0.84 &   0.30\\
  14 & 0.020 &  380 & 4635.85 & -5.750e+01 & -4.600e+01 &  6.402e-02 & -7.687e+00 & -7.701e-06 &  1.477e-02 & 4556 & 6374 &  -0.84 &   0.39\\
  15 & 0.018 &  395 & 4661.54 & -1.697e+01 &  2.606e+01 &  4.171e-02 &  3.297e+00 & -5.604e-06 &  3.134e-03 & 4578 & 6287 &  -0.78 &   0.39\\
  16 & 0.037 &  384 & 4690.14 & -1.012e+01 & -2.301e+01 &  4.402e-02 &  1.871e+00 & -5.633e-06 &  1.152e-02 & 4556 & 6403 &  -0.84 &   0.35\\
  17 & 0.021 &  165 & 4741.53 &  1.279e+02 & -5.548e+01 &  3.355e-03 & -7.231e+00 & -2.223e-06 &  1.614e-02 & 5107 & 6374 &  -0.83 &   0.21\\
  18 & 0.024 &  365 & 4749.95 & -1.323e+02 &  2.246e+01 &  8.116e-02 &  2.547e+00 & -8.989e-06 &  3.883e-03 & 4578 & 6403 &  -0.72 &   0.39\\
  19 & 0.024 &  356 & 4757.58 & -7.143e+01 & -2.808e+01 &  6.850e-02 & -1.834e+00 & -8.017e-06 &  1.250e-02 & 4578 & 6374 &  -0.84 &   0.35\\
  20 & 0.024 &  387 & 4799.41 & -1.553e+02 & -3.570e+01 &  8.893e-02 & -4.295e+00 & -9.673e-06 &  1.326e-02 & 4578 & 6403 &  -0.78 &   0.35\\
  21 & 0.020 &  374 & 4802.88 & -9.156e+01 & -3.493e+01 &  7.265e-02 & -3.840e+00 & -8.020e-06 &  1.323e-02 & 4556 & 6374 &  -0.84 &   0.33\\
  22 & 0.025 &  331 & 4808.15 & -5.913e+01 & -1.793e+01 &  5.758e-02 &  3.759e+00 & -7.360e-06 &  1.016e-02 & 4556 & 6289 &  -0.68 &   0.39\\
  23 & 0.017 &  237 & 4809.94 &  1.115e+02 &  6.798e+01 & -8.873e-03 &  1.471e+01 & -1.222e-06 & -5.567e-03 & 4556 & 6161 &  -0.56 &   0.39\\
  24 & 0.028 &  210 & 4811.05 &  5.310e+02 &  1.336e+02 & -1.522e-01 &  1.884e+01 &  1.087e-05 & -1.721e-02 & 4556 & 5917 &  -0.56 &   0.39\\
  25 & 0.018 &  215 & 4885.43 &  2.976e+01 & -3.157e+01 &  3.593e-02 & -9.299e+00 & -4.990e-06 &  1.183e-02 & 4679 & 6403 &  -0.84 &   0.30\\
  26 & 0.027 &  385 & 4952.65 & -3.128e+02 & -2.876e+00 &  1.446e-01 &  4.440e+00 & -1.400e-05 &  1.007e-02 & 4578 & 6374 &  -0.84 &   0.33\\
  27 & 0.024 &  340 & 4961.92 &  3.544e+02 &  8.169e+01 & -8.764e-02 &  1.032e+01 &  5.360e-06 & -6.613e-03 & 4647 & 6361 &  -0.70 &   0.39\\
  28 & 0.039 &   94 & 4967.90 & -6.541e+02 & -2.072e+02 &  2.631e-01 & -1.695e+01 & -2.356e-05 &  4.177e-02 & 4679 & 6374 &  -0.84 &   0.04\\
  29 & 0.018 &  249 & 4993.70 &  5.145e+02 &  1.031e+02 & -1.158e-01 &  1.718e+01 &  6.340e-06 & -5.782e-03 & 4989 & 6403 &  -0.84 &   0.30\\
  30 & 0.033 &  394 & 5054.65 &  1.340e+02 &  1.432e+00 & -3.989e-03 &  1.245e+01 & -2.134e-06 &  8.187e-03 & 4556 & 6403 &  -0.68 &   0.35\\
  31 & 0.014 &  156 & 5109.65 &  4.803e+02 &  3.256e+01 & -1.157e-01 &  3.659e-01 &  7.850e-06 &  3.189e-03 & 4999 & 6374 &  -0.83 &   0.18\\
  32 & 0.019 &  358 & 5223.19 &  5.883e+01 &  2.439e+01 &  1.539e-02 &  4.940e+00 & -3.539e-06 &  3.085e-03 & 4556 & 6361 &  -0.70 &   0.39\\
  33 & 0.023 &  158 & 5225.53 &  3.136e+01 & -2.554e+01 &  5.535e-02 & -1.776e+00 & -8.344e-06 &  1.168e-02 & 5023 & 6403 &  -0.84 &   0.30\\
  34 & 0.032 &  364 & 5228.38 & -1.692e+02 & -3.751e+01 &  1.017e-01 & -8.345e+00 & -1.073e-05 &  1.447e-02 & 4578 & 6403 &  -0.84 &   0.33\\
  35 & 0.020 &  289 & 5243.78 & -4.838e+01 & -1.845e+01 &  6.173e-02 & -2.325e+00 & -7.368e-06 &  1.151e-02 & 4647 & 6374 &  -0.84 &   0.30\\
  36 & 0.018 &  186 & 5247.06 &  1.322e+02 & -2.566e+01 &  2.116e-02 & -2.135e+00 & -5.616e-06 &  1.181e-02 & 5023 & 6361 &  -0.84 &   0.30\\
  37 & 0.037 &  194 & 5250.21 & -2.069e+02 &  7.157e+00 &  1.410e-01 &  3.865e+00 & -1.618e-05 &  6.933e-03 & 5009 & 6403 &  -0.84 &   0.30\\
  38 & 0.030 &  345 & 5288.53 & -1.348e+02 & -3.033e+01 &  8.871e-02 & -3.576e+00 & -9.581e-06 &  1.314e-02 & 4556 & 6374 &  -0.84 &   0.31\\
  39 & 0.018 &  355 & 5295.32 & -1.179e+02 &  2.315e+01 &  7.142e-02 &  4.884e+00 & -7.923e-06 &  3.454e-03 & 4578 & 6361 &  -0.56 &   0.39\\
  40 & 0.023 &  137 & 5376.83 & -1.145e+02 &  2.713e+01 &  6.600e-02 &  1.654e+01 & -7.583e-06 &  6.292e-04 & 4556 & 5914 &  -0.35 &   0.39\\
  41 & 0.016 &  334 & 5379.58 & -7.200e+00 & -1.588e+00 &  4.834e-02 & -1.483e+00 & -6.308e-06 &  8.373e-03 & 4624 & 6374 &  -0.84 &   0.33\\
  42 & 0.025 &  379 & 5386.34 & -4.134e+00 &  1.323e+01 &  3.585e-02 & -3.517e-01 & -5.080e-06 &  5.200e-03 & 4578 & 6374 &  -0.72 &   0.35\\
  43 & 0.022 &  229 & 5395.22 & -1.735e+02 &  5.967e+01 &  8.896e-02 &  2.075e+01 & -9.590e-06 & -4.265e-03 & 4556 & 6161 &  -0.38 &   0.39\\
  44 & 0.030 &  205 & 5398.28 & -2.415e+02 & -1.068e+02 &  1.248e-01 & -8.674e+00 & -1.222e-05 &  2.583e-02 & 4649 & 6374 &  -0.84 &   0.21\\
  45 & 0.026 &  375 & 5406.78 &  1.048e+02 &  2.616e+01 & -6.547e-04 &  3.715e+00 & -1.954e-06 &  3.123e-03 & 4647 & 6403 &  -0.78 &   0.39\\
  46 & 0.023 &  363 & 5417.04 & -9.121e+01 &  3.169e+01 &  6.766e-02 &  1.222e+01 & -7.924e-06 &  3.116e-03 & 4556 & 6361 &  -0.68 &   0.39\\
  47 & 0.038 &  211 & 5432.95 &  8.162e+01 & -5.922e+01 &  7.810e-03 & -4.982e+00 & -1.711e-06 &  1.828e-02 & 4578 & 6374 &  -0.84 &   0.19\\
  48 & 0.015 &  399 & 5436.30 & -9.556e+01 &  2.742e+01 &  7.104e-02 &  6.297e+00 & -8.221e-06 &  3.754e-03 & 4578 & 6374 &  -0.78 &   0.39\\
  49 & 0.026 &  383 & 5436.59 & -3.519e+01 & -1.176e+01 &  6.364e-02 &  5.610e+00 & -8.628e-06 &  1.091e-02 & 4556 & 6374 &  -0.78 &   0.39\\
  50 & 0.025 &  345 & 5441.34 &  3.332e+00 &  1.252e+02 &  3.525e-02 &  1.832e+01 & -5.217e-06 & -1.337e-02 & 4671 & 6374 &  -0.60 &   0.39\\
  51 & 0.033 &  305 & 5461.55 &  4.943e+02 &  8.377e+01 & -1.402e-01 &  8.728e+00 &  1.023e-05 & -7.197e-03 & 4723 & 6361 &  -0.62 &   0.39\\
  52 & 0.019 &  387 & 5464.28 & -2.130e+02 & -2.285e+01 &  1.091e-01 & -1.938e+00 & -1.135e-05 &  1.157e-02 & 4556 & 6374 &  -0.78 &   0.39\\
  53 & 0.026 &  369 & 5466.99 & -8.944e+01 &  4.082e+01 &  6.822e-02 &  1.549e+01 & -8.064e-06 &  1.694e-03 & 4578 & 6374 &  -0.72 &   0.39\\
  54 & 0.030 &  204 & 5473.17 &  3.299e+01 &  7.443e+01 &  1.203e-02 &  1.686e+01 & -2.516e-06 & -7.635e-03 & 4556 & 5973 &  -0.41 &   0.39\\
  55 & 0.023 &  301 & 5481.25 & -1.329e+02 &  3.132e+01 &  1.024e-01 & -2.019e-01 & -1.194e-05 &  4.071e-03 & 4698 & 6287 &  -0.84 &   0.33\\
  56 & 0.023 &   97 & 5491.83 & -3.001e+02 &  8.196e+01 &  1.321e-01 &  3.658e+01 & -1.349e-05 & -9.679e-03 & 4723 & 5914 &  -0.11 &   0.39\\
  57 & 0.023 &  384 & 5522.45 &  2.861e+01 &  9.286e+00 &  2.923e-02 &  2.292e+00 & -4.600e-06 &  6.703e-03 & 4671 & 6374 &  -0.78 &   0.39\\
  58 & 0.035 &  360 & 5538.52 &  3.771e+01 &  3.565e-01 &  2.561e-02 &  1.713e+00 & -4.423e-06 &  7.891e-03 & 4624 & 6403 &  -0.72 &   0.39\\
  59 & 0.029 &  370 & 5546.51 &  2.280e+02 &  2.684e+01 & -3.851e-02 &  8.539e-01 &  1.358e-06 &  3.512e-03 & 4730 & 6374 &  -0.84 &   0.35\\
  60 & 0.025 &  384 & 5553.58 &  1.295e+02 &  2.011e+01 & -7.179e-03 & -2.188e+00 & -1.418e-06 &  4.183e-03 & 4578 & 6361 &  -0.78 &   0.39\\
  61 & 0.020 &  389 & 5560.22 & -1.011e+02 & -1.202e+00 &  7.554e-02 & -3.020e+00 & -8.479e-06 &  8.282e-03 & 4671 & 6374 &  -0.84 &   0.35\\
  62 & 0.023 &  373 & 5587.58 & -2.720e+02 & -2.587e+01 &  1.295e-01 &  3.830e+00 & -1.321e-05 &  1.214e-02 & 4624 & 6374 &  -0.78 &   0.39\\
  63 & 0.024 &  397 & 5618.64 & -9.415e+01 &  6.043e+00 &  7.345e-02 & -3.181e+00 & -8.369e-06 &  6.800e-03 & 4556 & 6374 &  -0.84 &   0.39\\
  64 & 0.024 &  355 & 5619.60 &  1.379e+02 &  7.949e+01 & -1.050e-02 &  1.627e+01 & -1.276e-06 & -5.066e-03 & 4556 & 6374 &  -0.56 &   0.39\\
  65 & 0.023 &  381 & 5635.83 & -5.765e+01 &  2.943e+01 &  5.610e-02 &  4.510e-01 & -6.946e-06 &  2.472e-03 & 4578 & 6374 &  -0.70 &   0.39\\
  66 & 0.019 &  242 & 5636.70 & -2.439e+02 &  9.714e+00 &  1.189e-01 &  5.799e+00 & -1.271e-05 &  4.970e-03 & 4556 & 6161 &  -0.43 &   0.39\\
  67 & 0.012 &  132 & 5638.27 &  1.676e+02 &  5.739e+00 & -5.661e-03 & -9.538e-01 & -1.721e-06 &  7.890e-03 & 4879 & 6374 &  -0.83 &   0.18\\
  68 & 0.017 &  257 & 5641.44 & -2.810e+01 &  3.885e+01 &  6.422e-02 &  5.051e-01 & -8.293e-06 &  3.287e-03 & 4578 & 6374 &  -0.84 &   0.30\\
  69 & 0.022 &  361 & 5649.99 & -8.049e+00 &  6.051e+01 &  3.329e-02 &  1.014e+01 & -4.440e-06 & -2.451e-03 & 4785 & 6374 &  -0.60 &   0.39\\
  70 & 0.019 &  216 & 5651.47 & -7.249e+01 &  4.275e+01 &  5.219e-02 &  8.845e+00 & -6.298e-06 & -1.049e-03 & 4578 & 6161 &  -0.41 &   0.39\\
  71 & 0.040 &  339 & 5652.32 & -1.100e+01 &  9.628e+01 &  3.323e-02 &  2.182e+01 & -4.610e-06 & -1.017e-02 & 4556 & 6361 &  -0.62 &   0.39\\
  72 & 0.026 &  272 & 5661.35 & -3.620e+00 &  1.124e+02 &  3.146e-02 &  2.313e+01 & -4.637e-06 & -1.280e-02 & 4556 & 6287 &  -0.54 &   0.39\\
  73 & 0.029 &  340 & 5667.52 & -5.022e+01 &  3.480e+01 &  6.549e-02 &  8.794e+00 & -8.263e-06 &  4.560e-03 & 4556 & 6374 &  -0.84 &   0.33\\
  74 & 0.017 &  315 & 5679.03 &  1.396e+02 &  1.971e+01 & -3.645e-03 & -4.596e+00 & -1.765e-06 &  5.075e-03 & 4749 & 6374 &  -0.84 &   0.33\\
  75 & 0.017 &  181 & 5715.09 &  1.853e+02 &  2.684e+01 & -7.341e-03 &  4.650e+00 & -2.123e-06 &  5.455e-03 & 4649 & 6403 &  -0.83 &   0.21\\
  76 & 0.016 &  350 & 5731.77 &  5.208e+01 &  7.093e+00 &  2.631e-02 & -1.336e+00 & -4.375e-06 &  7.272e-03 & 4649 & 6403 &  -0.84 &   0.33\\
  77 & 0.025 &  123 & 5738.24 & -1.048e+02 &  1.380e+02 &  6.458e-02 &  2.746e+01 & -7.641e-06 & -1.874e-02 & 4556 & 5914 &  -0.20 &   0.39\\
  78 & 0.020 &  362 & 5741.85 & -2.092e+01 &  4.235e+01 &  4.233e-02 &  6.322e+00 & -5.725e-06 &  3.309e-04 & 4578 & 6361 &  -0.70 &   0.39\\
  79 & 0.022 &  378 & 5752.04 & -8.798e+01 & -1.295e+01 &  7.141e-02 & -2.124e+00 & -8.080e-06 &  1.065e-02 & 4730 & 6374 &  -0.84 &   0.35\\
  80 & 0.018 &  327 & 5775.08 & -2.884e+01 & -2.385e+00 &  5.526e-02 & -2.220e+00 & -6.923e-06 &  8.942e-03 & 4728 & 6374 &  -0.84 &   0.33\\
  81 & 0.018 &  287 & 5778.46 &  1.890e+01 &  4.506e+01 &  3.776e-02 &  7.088e+00 & -6.442e-06 & -2.017e-04 & 4556 & 6161 &  -0.70 &   0.39\\
  82 & 0.030 &  387 & 5793.92 &  5.060e+01 &  5.856e+01 &  1.629e-02 & -2.570e-01 & -3.303e-06 & -3.114e-03 & 4649 & 6374 &  -0.72 &   0.39\\
  83 & 0.031 &  371 & 5806.73 & -2.200e+02 &  2.733e+01 &  1.206e-01 &  1.304e+00 & -1.263e-05 &  4.599e-03 & 4578 & 6374 &  -0.84 &   0.34\\
  84 & 0.032 &  356 & 5809.22 &  2.561e+02 &  1.174e+02 & -3.839e-02 &  7.787e+00 &  4.498e-07 & -1.101e-02 & 4578 & 6374 &  -0.84 &   0.35\\
  85 & 0.014 &  271 & 5814.81 & -5.541e+01 &  5.903e+01 &  4.968e-02 &  1.129e+01 & -6.253e-06 & -3.345e-03 & 4556 & 6227 &  -0.54 &   0.39\\
  86 & 0.029 &   99 & 5827.88 &  2.484e+02 &  1.757e+02 & -6.769e-02 &  2.119e+01 &  4.691e-06 & -2.744e-02 & 4595 & 5886 &  -0.32 &   0.39\\
  87 & 0.021 &  381 & 5852.22 &  5.740e+01 &  6.228e+01 &  1.907e-02 &  5.890e+00 & -3.794e-06 & -2.186e-03 & 4624 & 6403 &  -0.78 &   0.39\\
  88 & 0.015 &  266 & 5855.08 & -1.581e+02 &  4.554e+01 &  8.292e-02 &  1.032e+01 & -8.931e-06 & -1.120e-03 & 4578 & 6227 &  -0.38 &   0.39\\
  89 & 0.017 &  386 & 5856.09 & -1.141e+02 &  2.988e+01 &  7.359e-02 &  6.299e+00 & -8.302e-06 &  2.511e-03 & 4624 & 6403 &  -0.70 &   0.39\\
  90 & 0.035 &   95 & 5902.48 & -2.580e+02 &  4.483e+01 &  1.138e-01 &  2.863e+01 & -1.150e-05 & -4.066e-03 & 4723 & 5914 &  -0.07 &   0.39\\
  91 & 0.032 &  319 & 5916.26 &  1.395e+02 & -8.026e+00 &  5.336e-03 & -4.553e+00 & -3.450e-06 &  9.420e-03 & 4649 & 6374 &  -0.84 &   0.33\\
  92 & 0.030 &  383 & 5927.79 & -8.166e+01 & -7.428e+00 &  6.484e-02 & -2.926e+00 & -7.472e-06 &  9.326e-03 & 4578 & 6403 &  -0.78 &   0.39\\
  93 & 0.033 &  393 & 5929.68 & -6.996e+01 & -5.744e-02 &  5.875e-02 & -6.093e+00 & -6.873e-06 &  7.402e-03 & 4649 & 6403 &  -0.78 &   0.39\\
  94 & 0.022 &  136 & 5934.66 &  9.081e+01 & -1.922e+01 &  2.423e-02 & -4.132e+00 & -4.644e-06 &  1.236e-02 & 5023 & 6374 &  -0.84 &   0.21\\
  95 & 0.023 &  292 & 5956.70 &  1.979e+02 &  7.238e+00 & -4.371e-03 &  3.304e+00 & -3.624e-06 &  6.958e-03 & 4820 & 6403 &  -0.84 &   0.35\\
  96 & 0.019 &  305 & 6027.06 & -3.348e+01 & -1.738e+01 &  5.513e-02 & -1.116e+00 & -6.600e-06 &  1.110e-02 & 4671 & 6374 &  -0.84 &   0.30\\
  97 & 0.023 &  190 & 6056.01 & -1.434e+02 & -5.097e+01 &  9.439e-02 & -3.554e+00 & -9.882e-06 &  1.737e-02 & 4578 & 6374 &  -0.84 &   0.21\\
  98 & 0.016 &  142 & 6078.49 &  2.758e+02 &  6.500e+01 & -4.065e-02 &  3.100e+00 &  1.085e-06 & -1.103e-03 & 4879 & 6374 &  -0.84 &   0.18\\
  99 & 0.020 &  390 & 6079.01 & -7.753e+01 &  3.215e+01 &  6.533e-02 &  9.778e-01 & -7.603e-06 &  2.745e-03 & 4671 & 6403 &  -0.78 &   0.39\\
 100 & 0.017 &  374 & 6082.72 &  1.050e+02 &  4.141e+01 &  1.286e-02 &  5.201e+00 & -4.321e-06 &  7.475e-04 & 4624 & 6361 &  -0.84 &   0.39\\
 101 & 0.033 &  375 & 6089.57 & -4.388e+01 &  2.996e+01 &  5.058e-02 &  7.216e+00 & -6.371e-06 &  2.512e-03 & 4556 & 6361 &  -0.72 &   0.39\\
 102 & 0.020 &  220 & 6094.38 & -3.242e+02 &  3.546e-01 &  1.431e-01 &  7.829e+00 & -1.445e-05 &  7.125e-03 & 4556 & 6161 &  -0.28 &   0.39\\
 103 & 0.017 &  375 & 6096.67 &  3.028e+01 &  6.386e+01 &  3.210e-02 &  3.028e+00 & -5.310e-06 & -2.434e-03 & 4671 & 6361 &  -0.70 &   0.39\\
 104 & 0.026 &  165 & 6098.25 & -1.722e+02 &  4.669e+01 &  8.482e-02 &  2.465e+01 & -9.018e-06 & -3.001e-03 & 4556 & 6161 &  -0.25 &   0.39\\
 105 & 0.017 &  358 & 6151.62 &  8.572e+01 &  2.699e+00 &  2.460e-02 &  9.152e-01 & -5.321e-06 &  7.643e-03 & 4578 & 6374 &  -0.84 &   0.35\\
 106 & 0.031 &  325 & 6157.73 & -6.641e+01 & -2.480e+01 &  6.653e-02 & -8.756e+00 & -7.682e-06 &  1.155e-02 & 4578 & 6374 &  -0.84 &   0.35\\
 107 & 0.036 &  389 & 6165.36 & -1.978e+02 & -3.319e+01 &  1.046e-01 & -3.580e+00 & -1.081e-05 &  1.340e-02 & 4556 & 6374 &  -0.84 &   0.39\\
 108 & 0.015 &  192 & 6173.34 &  2.765e+02 &  2.456e+00 & -3.815e-02 & -1.829e+00 &  3.782e-07 &  7.731e-03 & 4879 & 6361 &  -0.83 &   0.30\\
 109 & 0.016 &  369 & 6187.99 &  1.513e+02 &  6.428e+01 & -4.207e-03 &  3.402e+00 & -2.372e-06 & -1.913e-03 & 4750 & 6403 &  -0.84 &   0.35\\
 110 & 0.015 &  159 & 6200.32 &  2.515e+02 & -1.202e+01 & -2.843e-02 & -2.718e+00 & -4.263e-07 &  1.051e-02 & 4900 & 6361 &  -0.83 &   0.21\\
 111 & 0.020 &  239 & 6220.79 &  9.946e+01 &  8.621e+01 & -2.926e-03 &  1.944e+01 & -1.916e-06 & -8.336e-03 & 4556 & 6161 &  -0.56 &   0.39\\
 112 & 0.018 &  336 & 6226.74 & -4.656e+01 &  6.611e+01 &  5.308e-02 &  1.105e+01 & -6.895e-06 & -3.410e-03 & 4624 & 6276 &  -0.56 &   0.39\\
 113 & 0.019 &  375 & 6229.24 & -1.655e+00 &  6.564e+00 &  4.449e-02 &  5.706e+00 & -6.467e-06 &  7.753e-03 & 4556 & 6374 &  -0.72 &   0.39\\
 114 & 0.038 &  322 & 6238.39 & -3.487e+02 & -3.509e+01 &  1.158e-01 &  1.462e+01 & -8.241e-06 &  1.371e-02 & 4831 & 6361 &  -0.67 &   0.39\\
 115 & 0.016 &  360 & 6240.65 &  6.372e+01 &  3.407e+00 &  3.232e-02 &  1.895e+00 & -6.022e-06 &  7.794e-03 & 4624 & 6403 &  -0.78 &   0.35\\
 116 & 0.016 &  375 & 6270.23 &  2.415e+01 &  1.398e+01 &  4.272e-02 &  2.571e+00 & -6.554e-06 &  5.993e-03 & 4624 & 6374 &  -0.84 &   0.35\\
 117 & 0.028 &  391 & 6315.81 & -1.179e+02 &  1.939e+01 &  7.977e-02 &  3.463e+00 & -9.014e-06 &  4.853e-03 & 4578 & 6403 &  -0.70 &   0.35\\
 118 & 0.017 &  140 & 6322.69 &  1.652e+02 & -8.515e+00 &  1.986e-03 & -3.437e+00 & -3.022e-06 &  9.911e-03 & 5023 & 6374 &  -0.84 &   0.19\\
 119 & 0.010 &   94 & 6358.68 &  1.589e+02 &  4.306e+01 &  3.146e-02 &  2.037e+00 & -7.674e-06 &  4.734e-03 & 5417 & 6374 &  -0.78 &   0.21\\
 120 & 0.036 &  397 & 6380.75 & -2.942e+02 & -2.097e+01 &  1.426e-01 & -5.316e+00 & -1.429e-05 &  1.144e-02 & 4556 & 6374 &  -0.84 &   0.35\\
 121 & 0.018 &  240 & 6392.54 & -1.515e+02 &  1.361e+01 &  9.765e-02 &  3.821e+00 & -1.188e-05 &  4.910e-03 & 4578 & 5989 &  -0.72 &   0.39\\
 122 & 0.024 &  233 & 6481.88 & -2.832e+01 & -4.473e+01 &  6.769e-02 &  2.801e-01 & -8.946e-06 &  1.604e-02 & 4649 & 6403 &  -0.83 &   0.30\\
 123 & 0.027 &  331 & 6498.94 &  2.621e+02 & -2.801e+01 & -3.054e-02 & -7.395e+00 & -1.192e-06 &  1.237e-02 & 4649 & 6374 &  -0.84 &   0.39\\
 124 & 0.021 &  247 & 6608.03 & -1.104e+02 &  3.988e+01 &  8.333e-02 &  5.617e+00 & -1.062e-05 &  1.996e-04 & 4578 & 5989 &  -0.72 &   0.39\\
 125 & 0.015 &  219 & 6609.12 &  1.566e+02 &  1.759e+01 &  4.808e-03 &  4.456e+00 & -3.536e-06 &  6.485e-03 & 4649 & 6374 &  -0.84 &   0.30\\
 126 & 0.023 &  229 & 6625.02 &  1.285e+02 &  6.232e+01 &  6.883e-03 &  8.233e+00 & -4.658e-06 & -3.701e-03 & 4556 & 5914 &  -0.72 &   0.39\\
 127 & 0.018 &  326 & 6627.55 & -1.703e+02 &  8.829e+01 &  9.358e-02 &  1.550e+01 & -1.023e-05 & -7.453e-03 & 4578 & 6260 &  -0.50 &   0.39\\
 128 & 0.016 &  380 & 6703.57 &  2.267e+01 &  2.445e+01 &  3.990e-02 &  3.450e+00 & -6.453e-06 &  4.010e-03 & 4556 & 6361 &  -0.78 &   0.39\\
 129 & 0.018 &  390 & 6705.11 & -8.592e+01 &  5.960e+01 &  7.263e-02 &  6.213e+00 & -8.585e-06 & -1.000e-03 & 4556 & 6361 &  -0.78 &   0.35\\
 130 & 0.022 &  229 & 6710.32 &  7.043e+01 &  7.075e+01 &  2.311e-02 &  1.057e+01 & -5.682e-06 & -4.916e-03 & 4556 & 5917 &  -0.72 &   0.39\\
 131 & 0.021 &  292 & 6713.05 & -2.097e+02 &  6.184e+01 &  1.048e-01 &  9.106e+00 & -1.111e-05 & -3.766e-03 & 4671 & 6161 &  -0.41 &   0.39\\
 132 & 0.017 &  236 & 6713.74 & -2.250e+02 &  3.797e+01 &  1.056e-01 &  1.325e+01 & -1.090e-05 & -3.997e-05 & 4647 & 6161 &  -0.32 &   0.39\\
 133 & 0.019 &  193 & 6725.36 & -2.724e+02 &  3.980e+01 &  1.287e-01 &  7.781e+00 & -1.362e-05 &  9.816e-05 & 4698 & 5989 &  -0.26 &   0.39\\
 134 & 0.020 &  382 & 6726.67 & -1.289e+01 &  4.785e+01 &  4.599e-02 &  2.696e+00 & -6.151e-06 &  5.169e-04 & 4671 & 6403 &  -0.78 &   0.39\\
 135 & 0.017 &  311 & 6733.15 & -2.182e+02 &  4.476e+01 &  1.088e-01 &  7.913e+00 & -1.148e-05 &  2.804e-05 & 4624 & 6227 &  -0.45 &   0.39\\
 136 & 0.019 &  172 & 6739.52 &  2.370e+01 &  6.136e+01 &  3.417e-02 &  9.389e+00 & -6.356e-06 & -4.288e-03 & 4556 & 5827 &  -0.72 &   0.39\\
 137 & 0.021 &  306 & 6786.86 & -1.150e+02 &  5.132e+01 &  7.364e-02 &  8.126e+00 & -8.511e-06 & -1.370e-03 & 4556 & 6276 &  -0.54 &   0.39\\
 138 & 0.024 &  110 & 6793.26 &  4.719e+01 &  2.173e+02 &  5.951e-04 &  5.214e+01 & -1.032e-06 & -3.675e-02 & 4647 & 5914 &  -0.20 &   0.39\\
 139 & 0.027 &  359 & 6806.85 & -4.387e+01 &  4.316e+01 &  6.488e-02 &  4.865e+00 & -8.845e-06 &  9.710e-04 & 4556 & 6361 &  -0.78 &   0.35\\
 140 & 0.021 &  376 & 6810.27 & -4.150e+01 &  3.646e+01 &  5.659e-02 &  1.603e+00 & -7.038e-06 &  2.640e-03 & 4595 & 6374 &  -0.84 &   0.35\\
 141 & 0.016 &  377 & 6820.37 & -2.306e+01 &  5.392e+01 &  4.775e-02 &  7.352e+00 & -6.336e-06 & -3.263e-04 & 4649 & 6374 &  -0.78 &   0.35\\
 142 & 0.027 &  367 & 6828.60 & -1.282e+02 &  1.418e+01 &  8.794e-02 &  1.754e+00 & -9.707e-06 &  6.572e-03 & 4556 & 6403 &  -0.84 &   0.33\\
 143 & 0.024 &  174 & 6837.01 & -1.395e+02 &  3.806e+01 &  7.098e-02 &  1.724e+01 & -7.564e-06 & -1.265e-03 & 4556 & 6161 &  -0.22 &   0.39\\
 144 & 0.028 &  345 & 6839.84 &  2.072e+02 &  1.071e+02 & -2.714e-02 &  1.279e+01 & -5.973e-07 & -1.090e-02 & 4556 & 6287 &  -0.78 &   0.39\\
 145 & 0.019 &  384 & 6842.69 & -7.707e+01 &  6.390e+01 &  6.450e-02 &  2.873e+00 & -7.684e-06 & -2.778e-03 & 4671 & 6289 &  -0.70 &   0.35\\
 146 & 0.025 &  328 & 6843.66 & -2.222e+02 &  3.112e-01 &  1.244e-01 & -1.969e+00 & -1.305e-05 &  9.038e-03 & 4595 & 6374 &  -0.84 &   0.33\\
 147 & 0.033 &  205 & 6855.72 &  1.443e+02 &  2.005e+02 & -3.124e-02 &  5.407e+01 &  1.691e-06 & -3.082e-02 & 4595 & 6161 &  -0.25 &   0.39\\
 148 & 0.017 &  272 & 6857.25 & -1.797e+02 &  1.675e+01 &  9.315e-02 &  8.259e+00 & -1.005e-05 &  4.228e-03 & 4556 & 6227 &  -0.52 &   0.39\\
 149 & 0.028 &  379 & 6858.15 & -1.407e+02 &  6.900e+00 &  9.054e-02 & -6.242e-01 & -9.878e-06 &  7.621e-03 & 4556 & 6374 &  -0.84 &   0.35\\
 150 & 0.025 &  244 & 6861.94 &  2.667e+02 &  9.333e+01 & -5.126e-02 &  2.030e+01 &  1.418e-06 & -8.969e-03 & 4578 & 6026 &  -0.72 &   0.39\\
 151 & 0.026 &  340 & 6862.50 & -2.833e+02 &  2.371e+01 &  1.330e-01 &  7.602e+00 & -1.359e-05 &  3.926e-03 & 4556 & 6361 &  -0.54 &   0.39\\

\end{longtable}
$\dagger$ The columns correspond to the following: $id$ is the ID number of the line; $ns$ is the number of stars used for line calibration (out of a total of 451); 
\textit{stddev} is the final standard deviation for each calibration; $\lambda$ is the wavelength of the calibrated line;  $c_0$, $c_1$, 
$c_2$, $c_3$, $c_4$, and $c_5$ are the fitting coefficients for each line; $T_{eff}^{min}$ and $T_{eff}^{max}$ are the limits in effective temperature of the 
stars used to calibrate each line; \textit{fehmin} and \textit{fehmax} are the limits in [Fe/H] of the stars used to calibrate each line.

}

\end{document}